\pdfoutput=1
\documentclass[a4paper,american,floatfix,pdftex,superscriptaddress,twoside,%
aip,jcp,
citeautoscript,%
reprint
]{revtex4-2}
\usepackage{amsfonts,amsmath,amssymb}
\usepackage[T1]{fontenc}
\usepackage[utf8]{inputenc}
\usepackage{graphicx}
\usepackage{hypernat}
\usepackage[ams,todos]{CMImacros}
\usepackage{siunitx}

\graphicspath{{./fig/}} 

\fxsetup{final}

\newlength{\figwidth}
\makeatletter%
\renewcommand*{\@fnsymbol}[1]{\ensuremath{\ifcase#1\or \|\or *\or **\or \mathparagraph\or
      \mathsection\or \dagger\or \ddagger\or \dagger\dagger \or \ddagger\ddagger \else\@ctrerr\fi}}
\makeatother

\newcommand{\cfeldesy}{\affiliation{Center for Free-Electron Laser Science CFEL, Deutsches
      Elektronen-Synchrotron DESY, Notkestrasse 85, 22607 Hamburg, Germany}}%
\newcommand{\uhhcui}{\affiliation{Center for Ultrafast Imaging, Universität Hamburg, Luruper
      Chaussee 149, 22761 Hamburg, Germany}}%
\newcommand{\uhhphys}{\affiliation{Department of Physics, Universität Hamburg, Luruper Chaussee 149,
      22761 Hamburg, Germany}}%
\newcommand{\granada}{\affiliation{Instituto Carlos I de F\'{\i}sica Te\'orica y Computacional and
      Departamento de F\'{\i}sica At\'omica, Molecular y Nuclear, Universidad de Granada, 18071
      Granada, Spain}}%
\newcommand{\ayemail}{\email[Email:~]{andrey.yachmenev@cfel.de}}%
\newcommand{\rgfemail}{\email[Email:~]{rogonzal@ugr.es}}%
\newcommand{\jkemail}{\email[Email:~]{jochen.kuepper@cfel.de}}%
\newcommand{\cmiweb}{\homepage[website:~]{https://www.controlled-molecule-imaging.org}}%

\begin{document}
\title{Molecular influence on nuclear-quadrupole-coupling effects in laser induced alignment}
\author{Linda V.\ Thesing}\cfeldesy\uhhcui\uhhphys%
\author{Andrey Yachmenev}\ayemail\cfeldesy\uhhcui%
\author{Rosario Gonz{\'a}lez-F{\'e}rez}\rgfemail\granada%
\author{Jochen Küpper}\jkemail\cmiweb\cfeldesy\uhhcui\uhhphys%
\date{\today}%
\begin{abstract}\noindent%
   We studied the effect of nuclear-quadrupole interactions on the field-free impulsive alignment of
   different asymmetric-top molecules. Our analysis is focused on the influence of the hyperfine-
   and rotational-energy-level structures. These depend on the number of nuclear spins, the
   rotational constants, and the symmetry of the tensors involved in the nuclear spin and external
   field interactions. Comparing the prototypical large-nuclear-spin molecules iodobenzene,
   1,2-diiodobenzene, 1,3-diiodobenzene, and 2,5-diiodobenzonitrile, we demonstrate that the
   magnitude of the hyperfine splittings compared to the rotational-energy splittings plays a
   crucial role in the spin-rotational dynamics after the laser pulse. Moreover, we point out that
   the impact of the quadrupole coupling on the rotational dynamics decreases when highly excited
   rotational states dominate the dynamics.
\end{abstract}
\maketitle

\section{Introduction}%
The coupling of rotational and nuclear spin angular momenta, known as hyperfine interaction, can
have a significant influence on the rotational dynamics of molecules and is at the core of the
nuclear magnetic resonance measurement~\cite{Altkorn:MP55:1, Yan:JCP98:6869, Cool:JCP103:3357,
   Zhang:JCP104:7027, Wouters:CP218:309, Sofikitis:JCP127:144307, Bartlett:PCCP11:142,
   Bartlett:PCCP12:15689, Grygoryeva:JCP147:013901}. There are several mechanisms of hyperfine
interaction, the most prominent is the nuclear-quadrupole coupling. This occurs in molecules
containing nuclei with partly filled shells, resulting in non-spherical nuclear shapes and the
appearance of an electric nuclear quadrupole moment. This quadrupole couples, electrostatically
interacts, with the field gradient induced by other moving charged particles in the molecule, \ie,
the nuclei and electrons.

Generally, nuclear-quadrupole coupling is strong in molecules containing heavy atoms, such as Se,
Br, I, Fe, Au, or Pt, which are also found in many biologically relevant molecules. Due to their
very large x-ray- and electron-scattering cross sections these atoms are often chemically attached
to molecular compounds and used as marker atoms in diffractive imaging
experiments~\cite{Spence:PTRSB369:20130309, Kuepper:PRL112:083002, Yang:Science361:64,
   Hunter:NatComm7:13388}. Furthermore, in Coulomb-explosion imaging these heavy atoms often show
good axial recoil enabling the observation of molecular orientation and internal structure during
dynamical processes~\cite{Barty:ARPC64:415, Christensen:PRL113:073005, Miller:ARPC65:583}.

An indispensable component of high-resolution imaging experiments is the laser-induced alignment of
molecules, enabling measurements to be performed in the molecule-fixed frame without orientation
averaging. To avoid perturbations due to the presence of external fields, alignment under field-free
conditions is often favorable. Field-free alignment relies on the preparation of rotational wave
packets by means of intense optical laser pulses~\cite{Stapelfeldt:RMP75:543}. Since the wave
packets are non-stationary states, they dephase and rephase periodically, changing the alignment
with time, even after the pulsed excitation. Typical pulse shapes range from the short kick
pulses~\cite{Seideman:PRL83:4971, RoscaPruna:JCP116:6579, Hamilton:PRA72:043402} to pulses with long
rising and short falling edges~\cite{Seideman:JCP115:5965, Underwood:PRL90:223001,
   Underwood:PRL94:143002, Goban:PRL101:013001, Chatterley:JCP148:221105}. The time-evolution of the
molecular alignment in field-free conditions is revival structure, ideally with nearly the same
degree of alignment appearing every rotational period. The latter is, however, only valid for
molecules with regular spacings between the rotational energy levels, such as rigid linear and
symmetric top molecules.

The revivals of molecular alignment are highly sensitive to the molecule's rotational energy level
structure. A detailed understanding of various molecular degrees of freedom that can couple to
rotations is thus needed to accurately predict the alignment dynamics of molecules. It is well
established that hyperfine interactions can have a significant impact on the rotational dynamics of
molecules prepared in single rotational states~\cite{Grygoryeva:JCP147:013901}. Furthermore,
experimental evidence of the influence of nuclear quadrupole effects on the impulsive alignment
dynamics was published recently for I$_2$ molecules~\cite{Thomas:PRL120:163202}. In computational
studies of linear and asymmetric top molecules, we found a nontrivial dependence of the nuclear
quadrupole effects on the intensity of the laser field~\cite{Yachmenev:JCP151:244118,
   Thesing:JPCA124:2225}.

Here, provide a deeper insight into the effect of the quadrupole coupling on the alignment for
different molecular species. We compared the results across asymmetric top molecules with different
number of nuclear quadrupolar nuclei, their different positions in the molecule, and different
molecular symmetries. Our molecular set contains C$_{2v}$-symmetric molecules with one and two
iodine atoms, \ie, iodobenzene and the 1,2- and 1,3-isomers of diiodobenzene, besides the already
investigated 1,4-isomer, and the C$_s$-symmetric molecules diiodobenzonitrile and
1,3-bromoiodobenzene. In addition, we consider a hypothetical modified single-spin
1,3-diiodobenzene. First, we summarize our theoretical approach in \autoref{sec:theory}. We then
present impulsive alignment results for low- (\autoref{sec:impulsive_alignment_1e11}) and
high-intensity (\autoref{sec:impulsive_alignment_1e12}) laser pulses for a rotational temperature of
$\Trot=0~\text{K}$. Finally, we analyze the rotational dynamics of cold thermal ensembles and
individual excited states.

\section{Theoretical Description}
\label{sec:theory}
\begin{figure}
   \includegraphics[width=\linewidth]{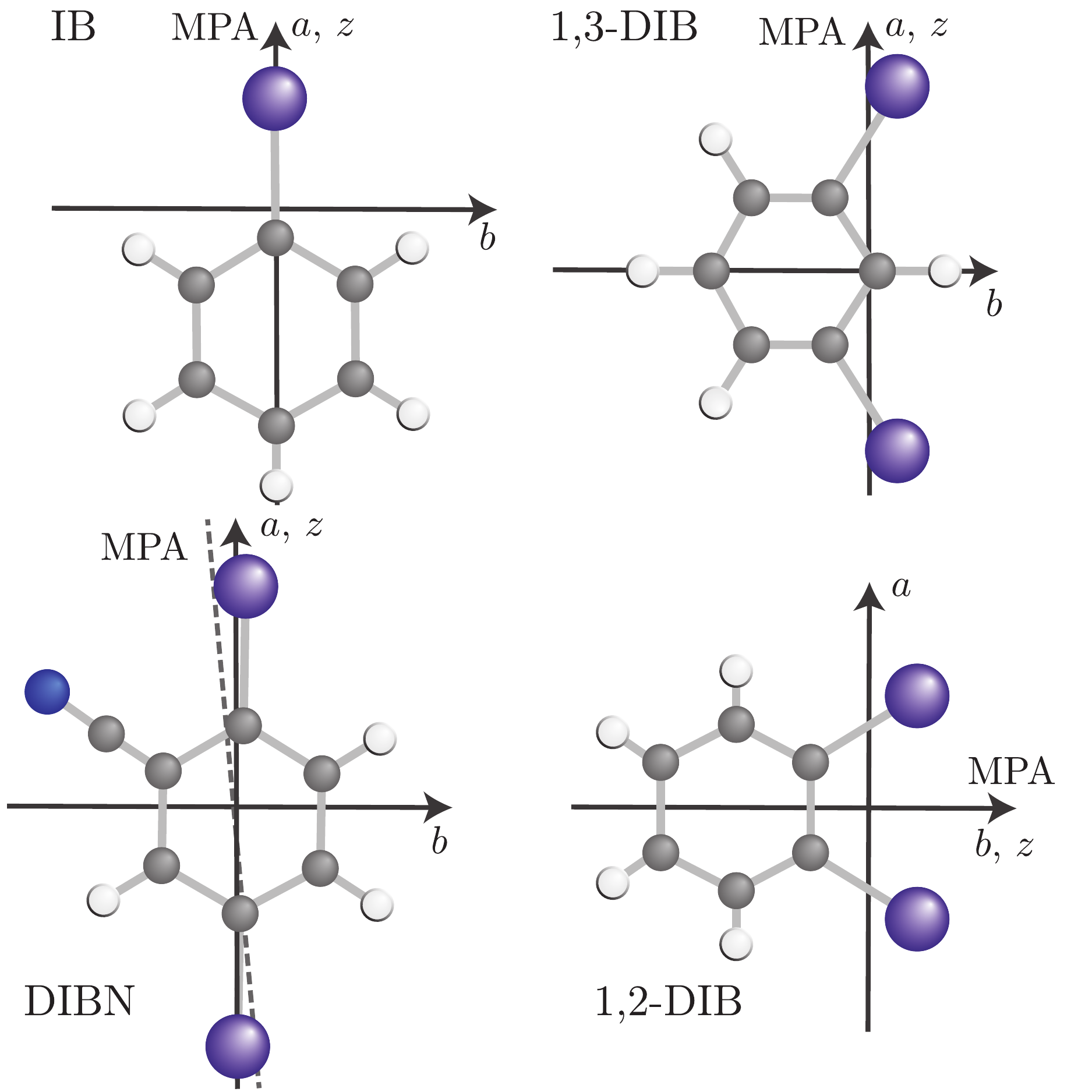}
   \caption{Sketches of the molecules IB, DIBN, 1,3-DIB and 1,2-DIB. For IB and 1,3-DIBN the most
      polarizable axis (MPA) is parallel to the $a$ axis, for DIBN it forms an angle of $\degree{5}$
      with the $a$ axis and for 1,2-DIB the MPA is parallel to the $b$ axis. The molecule-fixed $z$
      axis is indicated in the respective sketches.}
   \label{fig:molecule_sketches}
\end{figure}
We studied the laser-induced rotational dynamics of asymmetric top molecules taking into account the
nuclear-quadrupole interactions~\cite{Yachmenev:JCP151:244118}. The field-free spin-rotational
Hamiltonian of the system is given by
\begin{equation}
   \label{eq:hamiltonian_field_free}
   \hat{H}_\text{mol} = \hat{H}_\text{rot} +  \sum_{l} \mathbf{V}(l) \cdot \mathbf{Q}(l),
\end{equation}
where $\hat{H}_\text{rot}=A\hat{J}^2_a+B\hat{J}^2_b+C\hat{J}^2_c$ is the rigid-rotor Hamiltonian
with the rotational constants $A,B,C$ and the components $\hat{J}_i$, $i=a,b,c$, of the rotational
angular momentum operator $\hat{\mathbf{J}}$. The second term in the Hamiltonian
\eqref{eq:hamiltonian_field_free} describes the interaction of the quadrupole-moment tensor
$\mathbf{Q}(l)$ of the $l$-th nucleus with the electric-field-gradient (EFG) tensor $\mathbf{V}(l)$.
The sum runs over all nuclei with significant quadrupole moments $l$ in the molecule.

The interaction with the nonresonant laser field, linearly-polarized along the laboratory-fixed $Z$
axis, can be written as
\begin{equation}
   \hat{H}_\text{las} (t) = -\frac{I(t)}{2\varepsilon_0 c}\alpha_{ZZ}
   \label{eq:H_las}
\end{equation}
where $I(t)$ is the intensity of the laser field and $\alpha_{ZZ}$ is the corresponding element of
the molecule's polarizability tensor in the laboratory-fixed frame.

The geometrical structures, principal axes of inertia, and most-polarizable (MPA) axes of
iodobenzene (IB), diiodobenzonitrile (DIBN), 1,3-diiodobenzene (1,3-DIB) and 1,2-diiodobenzene
(1,2-DIB) are shown in \autoref{fig:molecule_sketches}. All four molecules contain one or two
$^{127}$I nuclei, which produce considerably large hyperfine energy splittings of the rotational
states due to iodine's large quadrupole moment $eQ=-696~\text{mb}$~\cite{Pyykko:MolPhys106:1965}.
The IB molecule has only one iodine nucleus and the total spin angular momentum is
$\hat{\mathbf{I}}=\hat{\mathbf{I}}_1$, where $I_1=5/2$. For the molecules with two iodine nuclei
$\hat{\mathbf{I}}=\hat{\mathbf{I}}_1+\hat{\mathbf{I}}_2$; the associated quantum number can have
values $I=0,...,5$. The total angular momentum is given by
$\hat{\mathbf{F}}=\hat{\mathbf{J}}+\hat{\mathbf{I}}$. The nitrogen nucleus in DIBN has an additional
nonzero quadrupole moment $eQ=20.44$~mb~\cite{Pyykko:MolPhys106:1965}, which is, however, relatively
small compared to iodine and thus neglected for the purpose of this study. This could also be
achieved by the $^{15}$N isotopologue with spin 1/2 and thus $eQ=0$. For 1,3-bromoiodobenzene
(1,3-BIB) we assume two different quadrupole-coupled nuclei $^{127}$I and $^{79}$Br, the more
abundant isotope of bromine with $Q=313~\text{mb}$~\cite{Pyykko:MolPhys106:1965}. All important
parameters for the molecules in this study, such as rotational constants, polarizability, and EFG
tensor elements are listed in \appautoref{sec:mol_param}.

The simulations of the spin-rotational dynamics of molecules subject to external electric fields
were performed using the variational approach Richmol~\cite{Owens:JCP148:124102,
   Yachmenev:JCP151:244118} and the same computational setup as described in our previous
work~\cite{Thesing:JPCA124:2225}. The total Hamiltonian is composed of the sum
$\hat{H}_\text{mol}+\hat{H}_\text{las}$ and the total time-dependent wave function is expanded in
the basis of eigenfunctions $\ket{F,\tilde{J}_{\tilde{K}_a\tilde{K}_c},n,M}$ of the field-free
time-independent Hamiltonian $\hat{H}_\text{mol}$. The details of the functional form of these basis
functions are described in \appautoref{sec:numerical_methods}. The alignment of the molecules is
quantified by the expectation value \costhreeD, where $\theta$ is the angle between the laser
polarization axis and the MPA of the molecule. In the present setup, the laser polarization is
aligned along the laboratory-fixed $Z$ axis and the MPA of all molecules is strictly or nearly
parallel to one of the principal axes of inertia, see \autoref{fig:molecule_sketches}. This
simplifies calculations of the expectation value \costhreeD, since the $\theta$ angle can be
associated with the corresponding Euler angle.

We compared the post-pulse molecular alignment with and without the nuclear-quadrupole-coupling term
in \eqref{eq:hamiltonian_field_free} for initial rotational temperatures
$\Trot=0,0.1,\text{~and~}0.3~\text{K}$. We used the eigenstates of $\hat{H}_\text{mol}$ and
$\hat{H}_\text{rot}$ as initial field-free molecular states in the calculations with and without
quadrupole coupling, respectively. We assumed equal populations for all hyperfine components of a
single rotational state, mimicking experimental conditions in molecular beams, where nuclear-spin
states typically do not cool and nuclear-spin temperatures correspond to the high-temperature limit
before expansion. Thus, even for rotational temperatures $\Trot=0~\text{K}$ the alignment was
calculated by averaging over the results of independent calculations for every hyperfine component
of the ground rotational state $J=0$. For the finite-temperature calculations further averaging was
performed by Boltzmann weighting the rotational states.

\section{Results and discussion}
\label{sec:results}
\subsection{Low-intensity impulsive alignment}
\label{sec:impulsive_alignment_1e11}
\begin{figure}
   \includegraphics[width=\linewidth]{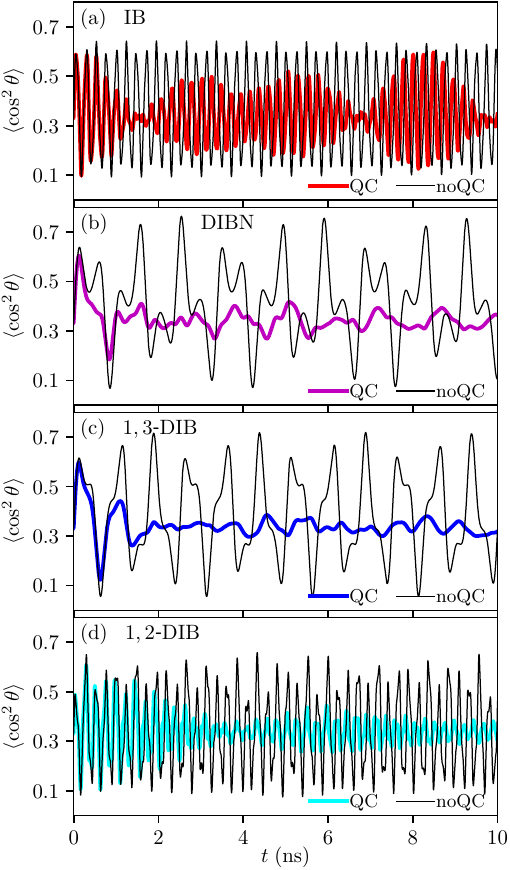}%
   \jknote{Remove parenthesis around sub-panel labels here and in all further figures.}
   \caption{Impulsive alignment induced by a Gaussian laser pulse with
      $\tau_\text{FWHM}=1~\text{ps}$ and $I_0=10^{11}~\Wpcmcm$ for (a) IB, (b) DIBN, (c) 1,3-DIB and
      (d) 1,2-DIB including (QC) and neglecting the nuclear-quadrupole coupling (noQC).}
   \label{fig:results_impulsive_1e11}
\end{figure}
Previously, we found that the effect of the nuclear-quadrupole coupling strongly depends on the
laser intensity and is larger at a lower intensity~\cite{Yachmenev:JCP151:244118,
   Thesing:JPCA124:2225}. Thus we begin by analyzing the influence of the quadrupole coupling on the
impulsive alignment induced by a 1~ps (FWHM) Gaussian laser pulse with a moderate peak intensity
$I_0 = 10^{11}~\Wpcmcm$. \autoref{fig:results_impulsive_1e11} shows the temporal evolution of the
post-pulse alignment dynamics, characterized by $\costhreeD(t)$, computed for different molecules at
$\Trot=0~\text{K}$. The results including nuclear-quadrupole coupling are plotted with thick lines
and the results of rigid-rotor calculations, \ie, neglecting the nuclear-quadrupole coupling, are
plotted with thin black lines.

In the rigid-rotor case, the alignment revival structures in \autoref{fig:results_impulsive_1e11}
look fairly regular for all molecules, which is the result of dephasing and rephasing of rather
narrow rotational wave packets. The revivals change from one molecule to another due to their
distinct rotational-energy-level structures and, to some extent, due to slightly different
interaction strengths with the laser field caused by the different polarizabilities. Due to the weak
intensity of the field, the post-pulse dynamics of IB, DIBN and 1,3-DIB is dominated by the
coherences between the low-energy rotational states with $K_a=0$. $K_a$ is the quantum number of the
rotational angular momentum projection operator onto the molecule-fixed $a$ axis. Since the energies
of these states look fairly similar to the energy levels of a symmetric top, the revival patterns in
\autoref[a--c]{fig:results_impulsive_1e11} repeat themselves after multiples of the respective
rotational periods $\taurot^\text{IB}=707.69~\text{ps}$, $\taurot^\text{DIBN}=3365.7~\text{ps}$ and
$\taurot^\text{1,3-DIB}=2499.4~\text{ps}$, with $\taurot=1/(B+C)$. In the case of 1,2-DIB, the
post-pulse dynamics (\autoref[d]{fig:results_impulsive_1e11}) does not show such a periodic
behavior, due to the molecule's larger rotational asymmetry, see \autoref{tab:parameters}.

The nuclear-quadrupole coupling increases the complexity of the post-pulse alignment dynamics. At
early times after the laser pulse \costhreeD does not show any significant deviation from the
rigid-rotor dynamics as the quadrupole coupling is much weaker than the polarizability interaction
with the laser field. However, the wavepackets quickly dephase~\cite{Thesing:JPCA124:2225},
characteristic of the non-regular hyperfine splittings of low-energy rotational states that dominate
the dynamics. Each rotational state of the IB molecule is split into 6 hyperfine components, at
most, while for the molecules with two iodine nuclei each state has up to 36 hyperfine components.
\begin{figure*}
   \includegraphics[width=\linewidth]{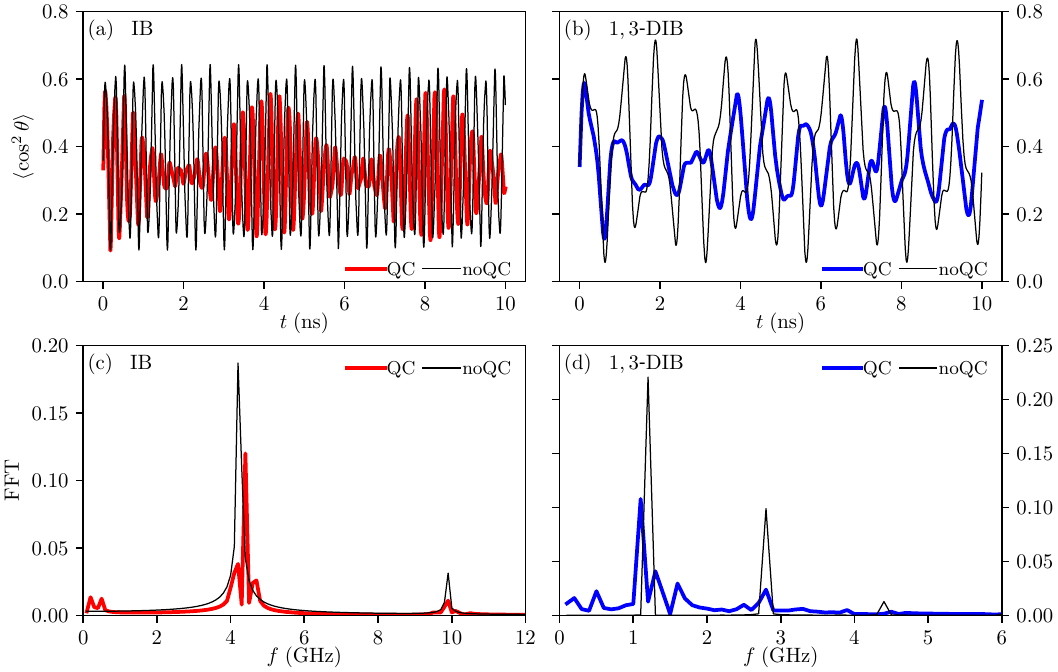}%
   \caption{Impulsive-alignment dynamics for $I_0=10^{11}~\Wpcmcm$ with nuclear-quadrupole coupling
      (QC) for the initial states (a) $\ket{5/2,0_{00},1/2}$ of IB and (b) $\ket{0,0_{00},0}$ of
      1,3-DIB and without the coupling (noQC). (c,~d) Fast-Fourier transforms of \costhreeD with and
      without quadrupole coupling.}
   \label{fig:fft_1e11}
\end{figure*}
The influence of the quadrupole coupling is noticeably stronger for DIBN and 1,3-DIB molecules, see
\autoref[b,~c]{fig:results_impulsive_1e11}. This is due to the rather small $B$ and $C$ rotational
constants for these molecules, corresponding rotational-level spacings in DIBN and 1,3-DIB at low
rotational excitations that are comparable to or even smaller than the hyperfine splittings. This
yields strong mixing of different rotational levels, even in the initial field-free states, giving
rise to large dephasing effects. A similar behavior was observed previously for
1,4-DIB~\cite{Thesing:JPCA124:2225}. For IB and 1,2-DIB, the hyperfine splittings are much smaller
than the rotational level spacings and thus the dephasing effect is weaker. Consequently, the
alignment for these molecules oscillates at frequencies similar to the rigid-rotor results, see
\autoref[a,~d]{fig:results_impulsive_1e11}. The effect of the quadrupole coupling is, however,
visible in the strongly modified amplitude of the degree of alignment.

In order to illustrate the interplay between the hyperfine-split and the pure rotational energy
level spacings, we computed the fast-Fourier transforms (FFTs) of $\costhreeD(t)$ for IB and 1,3-DIB
molecules, see \autoref{fig:fft_1e11}. Here, the wavepackets obtained from the single initial
hyperfine state $\ket{F,\tilde{J}_{\tilde{K}_a\tilde{K}_c},n,M}=\ket{5/2,0_{00},1,1/2}$ for IB or
$\ket{0,0_{00},1,0}$ for 1,3-DIB were analyzed.

The Fourier transforms of the rigid-rotor results show only a few quasi-regularly-spaced peaks,
reflecting the periodic behavior of \costhreeD. For both molecules, the highest intensity peak
corresponds to the coupling of the rigid-rotor states $\ket{J_{K_a,K_c}}=\ket{0_{00},0}$ and
$\ket{2_{02},0}$. When the quadrupole coupling is considered, the main rigid-rotor peaks are split
into several frequencies corresponding to transitions between different hyperfine states. For IB,
\autoref[c]{fig:fft_1e11}, the main peak at $f\approx4.2~\text{GHz}$ is shifted toward higher
frequency, which leads to a faster oscillation of the alignment than for the rigid-rotor, with the
opposite effect observed for the main peak $f\approx1.2~\text{GHz}$ in 1,3-DIB,
\autoref[d]{fig:fft_1e11}. The absolute value of the shift is similar for both molecules, however,
relative to the peak frequency the shift in 1,3-DIB is about four times bigger than in IB. Thus, for
1,3-DIB, the deviations between the phases in the post-pulse alignment for the result with and
without the quadrupole coupling occur on the timescale of the rotational revivals. In contrast for
IB, the effect is slower than the molecule's longer rotational period. Furthermore, the larger
number of hyperfine states in 1,3-DIB compared to IB led to an increased amount of irregularly
spaced frequency contributions, giving rise to the non-periodic dynamics, see
\autoref[b]{fig:fft_1e11}.

Other initial spin-rotational eigenstates of 1,3-DIB, which contribute to the alignment in
\autoref[c]{fig:results_impulsive_1e11}, show analogous patterns, but with different frequency
contributions. As a result, the revivals of \costhreeD are on average almost entirely suppressed in
1,3-DIB. For both molecules, the frequency components below $f\approx1~\text{GHz}$ originate from
the energy gaps between different hyperfine states belonging to the same rotational level, \ie, to
pure hyperfine-level population transfers similar to the transitions observed in NMR spectroscopy.

\subsubsection{Influence of the number and strengths of nuclear spins}
To gain more insight into the influence of the number of nuclear spins, we compared the alignment
dynamics of 1,3-DIB to that of a modified single-spin 1,3-DIB molecule, where we considered the
quadrupole coupling of only one iodine nucleus with all other properties of 1,3-DIB unchanged. This
allowed us to compare two molecular systems that differ only in the number of quadrupole-coupled
nuclei. Furthermore, we investigated the alignment of 1,3-BIB, which can be considered an
intermediate case between 1,3-DIB and the single-spin 1,3-DIB with respect to its rotational and
hyperfine splittings.

\begin{figure}[b]
   \includegraphics[width=\linewidth]{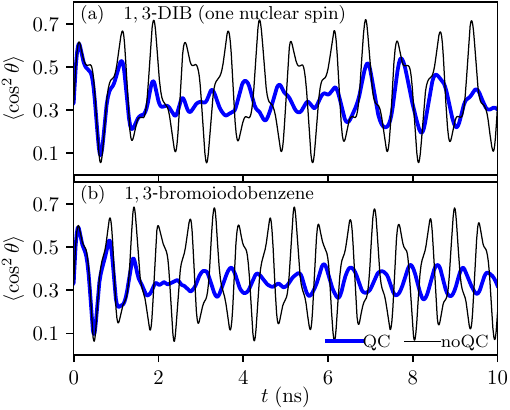}
   \caption{Impulsive alignment induced by $\tau_\text{FWHM}=1~\text{ps}$ laser pulses with
      $I_0=10^{11}~\Wpcmcm$ for (a) 1,3-DIB neglecting the quadrupole coupling of the second iodine
      nucleus and (b) 1,3-bromoiodobenzene including (QC) and neglecting the nuclear quadrupole
      coupling (noQC).}
   \label{fig:13DIB_modified}
\end{figure}
The alignment of the single-spin 1,3-DIB, shown in \autoref[a]{fig:13DIB_modified}, exhibits
stronger or longer-lasting revivals than the regular 1,3-DIB with two coupled iodine nuclei, see
\autoref[c]{fig:results_impulsive_1e11}.When comparing to a molecule with the same number of spins,
IB, the impact of the quadrupole coupling in the single-spin 1,3-DIB is still more pronounced,
\autoref[a]{fig:results_impulsive_1e11}, even though the quadrupole splittings for both molecules
have similar values. However, the pure rotational energy spacings in IB are much larger than in
1,3-DIB. This highlights that it is the relative differences between the hyperfine and the pure
rotational energy spacings in the molecule that determines the magnitude of the nuclear-quadrupole
effect in the alignment. This conclusion is further supported by the rotational dynamics of 1,2-DIB,
which has two iodine nuclei, but much larger rotational energy splittings than 1,3-DIB. As a result,
\costhreeD is affected by the quadrupole interactions in a manner that qualitatively resembles the
behavior of IB rather than 1,3-DIB. This can be seen, for example, when the peak alignment in
1,2-DIB increases after four nanoseconds in \autoref[d]{fig:results_impulsive_1e11}. The alignment
for 1,3-BIB with and without quadrupole coupling is depicted in \autoref[b]{fig:13DIB_modified}.
This molecule has fewer hyperfine energy levels and slightly larger rotational-energy spacings than
1,3-DIB. As a result, the low-amplitude oscillations of the alignment have a more regular structure
for 1,3-BIB, while the overall effect of the quadrupole coupling in 1,3-BIB is similar to that in
1,3-DIB, see \autoref[c]{fig:results_impulsive_1e11}.

\subsubsection{Influence of the geometric symmetry}
Apart from their rotational constants and the number of hyperfine levels, an important distinction
of the molecules considered here lies in the symmetry of their EFG and polarizability tensors. In IB
both tensors are diagonal in the coordinate system defined by the principal axes of inertia, while
the EFG tensors of the other molecules and the polarizability tensor of DIBN have nonzero
off-diagonal elements in the inertial frame. The diagonal elements of these tensors couple
rotational states having the same symmetry in the D$_2$ rotation group~\cite{Gordy:MWMolSpec}. The
off-diagonal elements $\alpha_{ab}$ and $V_{ab}$ couple rotational states with
$A\leftrightarrow{}B_c$ symmetry as well as $B_a\leftrightarrow{}B_b$. For DIBN, and the two
diiodobenzene molecules, however, a given spin-rotational eigenstate is a linear combination of
basis states $\ket{F,J_{K_aK_c},I,M}$ with $A$ and $B_c$ or $B_a$ and $B_b$ rotational symmetries.
For an eigenstate $\ket{F,\tilde{J}_{\tilde{K}_a\tilde{K}_c},n,M}$, only an approximate rotational
symmetry can be assigned according to even or odd $\tilde{K}_a$ and
$\tilde{K}_c$~\cite{Gordy:MWMolSpec}. Consequently, a weak laser-interaction coupling can be
observed between the hyperfine states corresponding to rotational states which are not coupled by
the laser field in the absence of the quadrupole interactions.

Without quadrupole interactions, the wave packets of IB, 1,3-DIB and 1,2-DIB consist exclusively of
rotational states of the same symmetry in the D$_2$ group as the initial state, \ie, $A$ symmetry
for $\Trot=0~\text{K}$. For DIBN, states with $B_c$ symmetry also contribute due to the
polarizability tensor's nonzero off-diagonal elements. When considering quadrupole coupling, nonzero
populations of $B_c$ rotational states are also observed for 1,3-DIB and 1,2-DIB. These
contributions are small, $\lesssim5~\%$, for the initial states and laser intensity considered here
and thus do not influence the alignment significantly. However, this aspect of the rotational
dynamics is not captured if the nuclear-quadrupole interactions are neglected. In addition, for
rotational levels which are strongly mixed by these interactions, a much larger impact on the
population distribution is possible. This is, for instance, the case for the excited states
$\ket{2_{12},M_J}$ and $\ket{2_{02},M_J}$ of 1,2-DIB, see \autoref{sec:thermal}.

\subsection{High-intensity impulsive alignment}
\label{sec:impulsive_alignment_1e12}
\begin{figure}[b]
   \includegraphics[width=\linewidth]{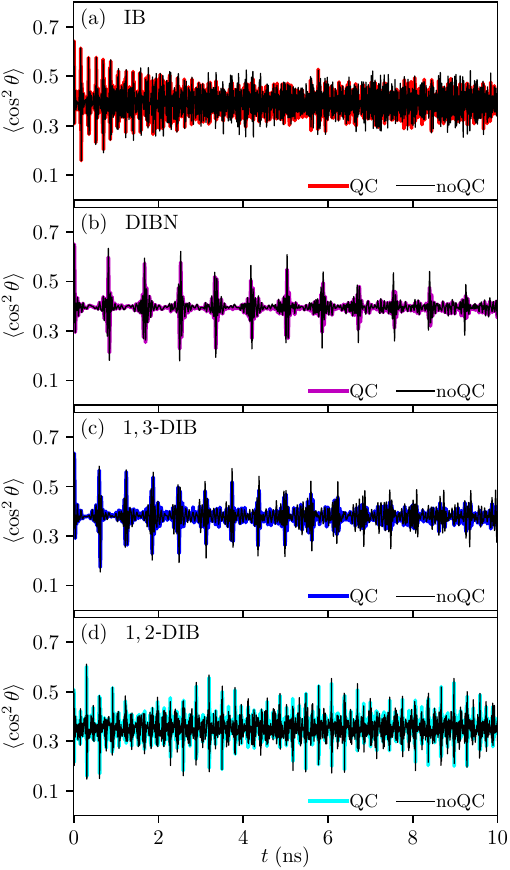}
   \caption{Impulsive alignment induced by a Gaussian laser pulse with
      $\tau_\text{FWHM}=1~\text{ps}$ and $I_0=10^{12}~\Wpcmcm$ for (a) IB, (b) DIBN, (c) 1,3-DIB and
      (d) 1,2-DIB including (QC) and neglecting the nuclear-quadrupole coupling (noQC).}
   \label{fig:results_impulsive_1e12}
\end{figure}
Typical impulsive alignment experiments employ intenser laser pulses~\cite{RoscaPruna:PRL87:153902,
   Holmegaard:PRA75:051403R, Trippel:PRA89:051401R, Karamatskos:NatComm10:3364}, which excite a
larger number of high-energy rotational states than discussed in the previous section. We
investigated the strong field alignment using the same pulse shape as in the previous section but
with the peak intensity increased tenfold, \ie, $I_0=10^{12}~\Wpcmcm$. The temporal evolution of
\costhreeD for all molecules except 1,2-DIB is characterized by $J$-type
revivals~\cite{Poulsen:JCP121:783, Holmegaard:PRA75:051403R} that appear at multiples of the
rotational periods as well as integer fractions thereof, see \autoref{fig:results_impulsive_1e12}.
These beatings originate from the interference of highly excited rotational states with
$\Delta{}K_a=0$. The asymmetry splittings of these states prevent a complete rephasing of the wave
packet and the peak alignment at the revivals decreases over time~\cite{Rouzee:PRA73:033418}.

For 1,2-DIB, we observe clear revival features of a different type spaced by $304~\text{ps}$ in
\autoref[d]{fig:results_impulsive_1e12}. The post-pulse dynamics of this molecule is dominated by
the dephasing and rephasing of rotational states with $\Delta{}J=\Delta{}K_a =2$ and
$J\approx{}K_a$. For large $J$, the energies of these states can be approximated by
$E(J,K_a)\approx\overline{B}J(J+1)+(A-\overline{B})K_a^2$ with $\overline{B}=\frac{1}{2}(B+C)$. The
relevant energy gaps are thus multiples of $4A$ yielding the period of $1/4A\approx304~\text{ps}$ in
\costhreeD.
\jknote{Name revivals – didn't these later ones also have a name in RCS? -- where the $J$-type
   revival name also comes from \pdfpar Linda: There are A-type and C-type revivals but from what I
   understand about their origin, it is not exactly the same (I could be wrong). \pdfpar Jochen:
   Rosario, see \cite{Felker:JPC96:7844,Felker:JPC90:724} and/or \cite{Riehn:CP283:297} -- both have
   tables with the ``revival types''.}

For the stronger laser pulse, the nuclear-quadrupole interactions had a much weaker impact on the
post-pulse rotational dynamics. generally, it led to a loss of the peak alignment over time instead
of qualitatively altering the dynamics~\cite{Thesing:JPCA124:2225}, see
\autoref{fig:results_impulsive_1e12}. An exception of this primarily destructive effect of the
quadrupole coupling is found at certain alignment peaks of 1,2-DIB, \eg, $t=5.17~\text{ns}$, where
\costhreeD with quadrupole coupling is slightly larger than without. We attribute this constructive
effect of the quadrupole coupling to the hyperfine splittings of some low-energy rotational states
that cause a change in the amplitude of \costhreeD similar to the weak laser intensity regime,
\autoref[c]{fig:results_impulsive_1e11}. Overall, the weak impact of the quadrupole interactions is
a result of increasingly uniform hyperfine splitting patterns at higher rotational excitations, as
previously shown for I$_2$ and 1,4-DIB~\cite{Thesing:JPCA124:2225}.

\begin{figure*}
   \includegraphics[width=\linewidth]{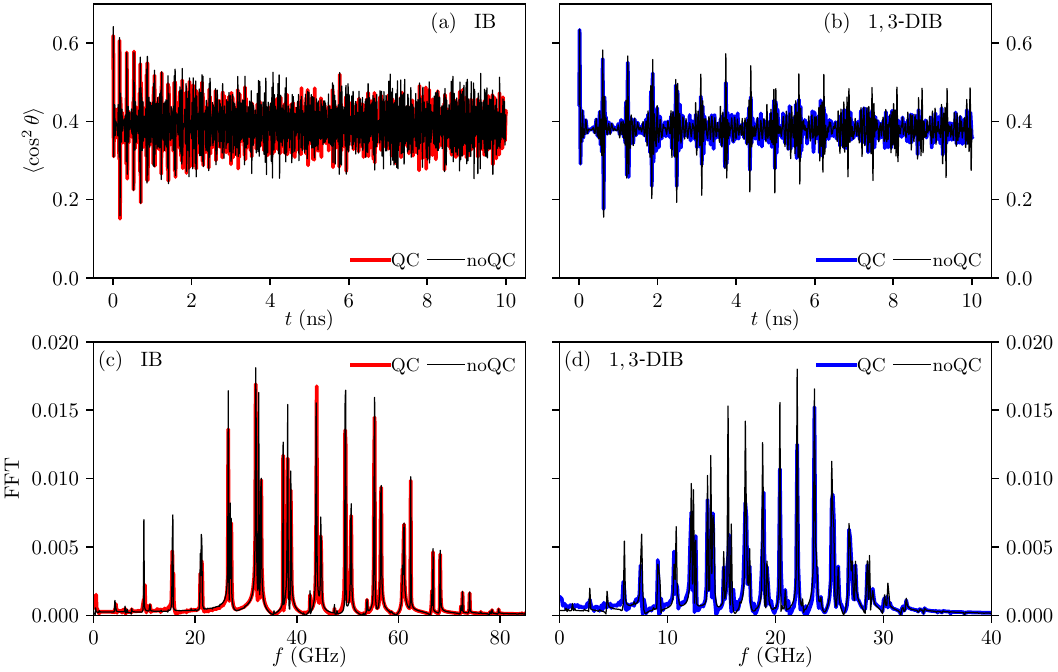}
   \caption{Impulsive alignment for $I_0=10^{12}~\Wpcmcm$ with (QC) and without (noQC)
      nuclear-quadrupole coupling for the initial states (a) $\ket{5/2,0_{00},1/2}$ of IB and (b)
      $\ket{0,0_{00},0}$ of 1,3-DIB. (c,~d) Fast-Fourier transforms of \costhreeD dynamics of the
      dynamics in (a) and (b).}
   \label{fig:fft_1e12}
\end{figure*}
The FFTs of the post-pulse alignment for IB and 1,3-DIB are shown in \autoref[c,~d]{fig:fft_1e12}.
The stronger-intensity pulse creates much broader wave packets that contain multiple frequency
components. For both molecules the Fourier transforms calculated with and without quadrupole
interactions are similar, especially for higher frequencies $f$ that correspond to transitions
between highly-excited rotational states. Deviations occur mainly in the lower frequency range,
which contributes to a slight dephasing and the corresponding decrease of the alignment at longer
timescales, noticed in \autoref[a,~b]{fig:fft_1e12}. As for the weak laser intensity, the dephasing
is more pronounced in 1,3-DIB due to the stronger mixing of less separated neighboring rotational
states and generally larger number of the hyperfine components.

\subsection{Influence of the temperature}
\label{sec:thermal}
\begin{figure*}
   \includegraphics[width=\linewidth]{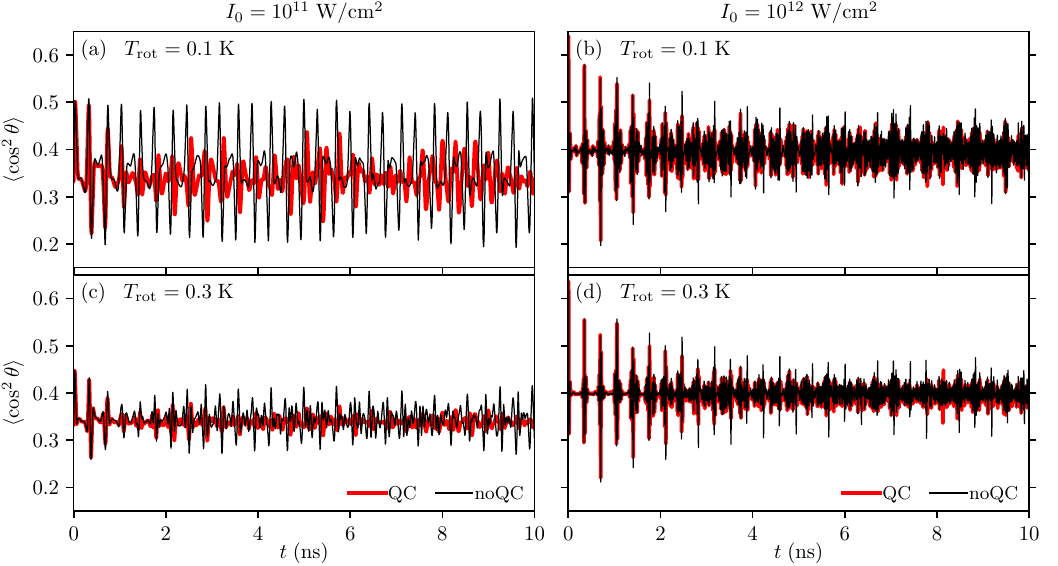}%
   \caption{Impulsive alignment of a thermal ensemble of IB molecules induced by laser pulses with
      $\tau_\text{FWHM}=1~\text{ps}$ including and neglecting the quadrupole coupling (QC and noQC)
      for (a,~b) $\Trot=0.1~\text{K}$ and (c,~d) $\Trot=0.3~\text{K}$. (a, c) show the results for
      $I_0=10^{11}~\Wpcmcm$ and (b, d) for $I_0=10^{12}~\Wpcmcm$.}
   \label{fig:IB_thermal}
\end{figure*}
While single-state molecular ensembles can be achieved for small molecules~\cite{Chang:IRPC34:557,
   Nielsen:PCCP13:18971, Horke:ACIE53:11965, Karamatskos:NatComm10:3364, Meerakker:CR112:4828}, this
is generally not feasible for larger molecules~\cite{Meerakker:CR112:4828, Wohlfart:PRA77:031404R,
   Bethlem:JPB39:R263}. We thus need to understand the effect of the nuclear-quadrupole coupling at
finite rotational temperatures. We computed the post-pulse impulsive alignment of thermal ensembles
of IB molecules for the two laser intensities used above and the rotational temperatures
$\Trot=0.1~\text{K}$ and $\Trot=0.3~\text{K}$. Such thermal ensembles are
comparable~\cite{Thesing:JCP146:244304} to state selected molecular
ensembles~\cite{Filsinger:PCCP13:2076, Meerakker:CR112:4828, Chang:IRPC34:557} and represent
conditions that are experimentally achievable.

The alignment of the thermal ensembles is shown in \autoref{fig:IB_thermal}. Without quadrupole
interactions, \costhreeD shows revival features with a decreased degree of alignment as compared to
$\Trot=0~\text{K}$, which is due to the influence of 32 (130) initially populated excited states
with up to $J=3$ ($J=5$) for $\Trot=0.1~\text{K}$ ($\Trot=0.3~\text{K}$). As for $\Trot=0~\text{K}$,
the quadrupole-coupling effects are stronger at the weaker field intensity $I_0=10^{11}~\Wpcmcm$,
see \autoref[a, c]{fig:IB_thermal}. For this laser intensity, a periodic increase of the amplitude
of \costhreeD as observed for $\Trot=0~\text{K}$ in \autoref[a]{fig:results_impulsive_1e11} is not
present. At a stronger intensity $I_0=10^{12}~\Wpcmcm$ the effect of the quadrupole coupling is
essentially negligible until the first full revival, see \autoref[b, d]{fig:IB_thermal}. Shortly
thereafter the quadrupole interaction starts to manifest itself, reducing the overall degree of
alignment.

\begin{figure}
   \includegraphics[width=\linewidth]{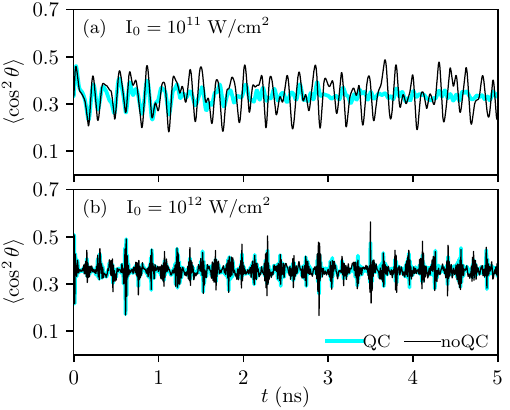}
   \caption{Impulsive alignment averaged over the initial states $\ket{2_{12},M_J}$ of 12-DIB
      induced by $1~\text{ps}$ laser pulses with (a) $I_0=10^{11}~\Wpcmcm$, (b)
      $I_0=10^{12}~\Wpcmcm$. The alignment is shown including (QC) and neglecting the nuclear
      quadrupole coupling (noQC).}
   \label{fig:12DIB_excited}
\end{figure}
To account for the nuclear-quadrupole effect in thermal ensembles of different molecular species, we
additionally computed the alignment dynamics of the excited rotational state $\ket{2_{12}}$ of
1,2-DIB, see \autoref{fig:12DIB_excited}. We compared the average rigid-rotor alignment over all
possible $M_J$ values of the initial state to the average result obtained for all initial hyperfine
states $\ket{F,\tilde{2}_{\tilde{1}\tilde{2}},n,M}$. As for the rotational ground state, the effect
of the quadrupole coupling is stronger for weaker intensity laser field. For $I_0=10^{12}~\Wpcmcm$
(\autoref[b]{fig:12DIB_excited}) the effect is still more pronounced than for $\Trot=0~\text{K}$ in
\autoref[d]{fig:results_impulsive_1e11}. This can be attributed to the nuclear-quadrupole
interactions strongly mixing the rotational states $\ket{2_{12}}$ and $\ket{2_{02}}$, which are not
coupled by the polarizability interaction. This mixing led to significant contributions of basis
states $\ket{F,2_{02},I,M}$ in the initial wave functions
$\ket{F,\tilde{2}_{\tilde{1}\tilde{2}},n,M}$ modifying the dynamics even before the laser field was
applied. After the interaction with the laser pulse, we find significant populations of hyperfine
states belonging to rotational states that are otherwise not excited when the quadrupole coupling is
neglected.

We thus conclude that for finite-temperature molecular beams the influence of the nuclear-quadrupole
interaction generally decreases with increasing the laser intensity. This is further strengthened by
our previous analysis of the excited state rotational dynamics of I$_2$ and
1,4-DIB~\cite{Thesing:JPCA124:2225}. However, the hyperfine interactions can have a significant
impact on the dynamics of some excited initial states and cannot be neglected even on a time scale
where they do not affect the alignment for $\Trot=0~\text{K}$.

\section{Summary and Conclusions}
We presented a computational analysis of the post-pulse alignment dynamics of asymmetric top
molecules with strong nuclear-quadrupole interactions. We focused on differences in the hyperfine
interactions due to different chemical structures of the molecules. We showed that the interplay of
hyperfine energy shifts and rotational energy level spacings is a deciding factor for the extent to
which the coupling affects the post-pulse alignment. For all molecules analyzed in the present work
and in previous studies~\cite{Yachmenev:JCP151:244118, Thesing:JPCA124:2225} we observed that the
influence of the quadrupole coupling becomes small when the field-dressed dynamics is dominated by
excited rotational states. Due to the variety of molecules considered, we expect that this
observation is general and that pushing population into highly-excited rotational states is
advantageous for minimizing the dephasing due to nuclear-quadrupole coupling.

We point out that while nuclear-quadrupole interactions pose difficulties in achieving strong
field-free alignment, their effect on the rotational dynamics might be exploited to probe nuclear or
electronic excitations which alter the nuclear spins and quadrupole momenta as well as electric
field gradients. The sensitivity of the alignment to the hyperfine energy level structures of
low-energy rotational states may thus be viewed as an opportunity rather than a disadvantage.

\section{Acknowledgements}
This work was supported by Deutsches Elektronen-Synchtrotron DESY, a member of the Helmholtz
Association (HGF), including the maxwell compute-cluster operated by DESY, by the Deutsche
Forschungsgemeinschaft (DFG) through the priority program ``Quantum Dynamics in Tailored Intense
Fields'' (QUTIF, SPP1840, KU~1527/3, YA~610/1), and through the clusters of excellence ``Center for
Ultrafast Imaging'' (CUI, EXC~1074, ID~194651731) and ``Advanced Imaging of Matter'' (AIM, EXC~2056,
ID~390715994). R.G.F.\ gratefully acknowledges financial support by the Spanish Project
PID2020-113390GB-I00 (MICIN), and by the Andalusian research group FQM-207.

\appendix
\section{Molecular parameters}
\label{sec:mol_param}
\begin{table*}
   \centering
   \renewcommand{\arraystretch}{1.2}
   \begin{tabular}{l | S[table-format=7.7] | S[table-format=5.5] | S[table-format=5.5] | S[table-format=5.5] | S[table-format=5.5]}
     \hline \hline
     & {IB} $^\text{a}$ & {DIBN} & {1,3-DIB} & {1,2-DIB} & {1,3-BIB}\\
     \hline
     $A$ (MHz) & 5669.131 & 1608.324 & 1859.912 & 822.0850 & 2040.244 \\
     $B$ (MHz) & 750.414293 & 155.402 & 210.776 & 522.1492 & 281.807 \\
     $C$ (MHz) & 662.636146 & 141.709 & 189.321 & 319.3276 & 247.607 \\
     $\kappa$  & -0.965 & -0.981 & -0.974 & -0.193 & -0.962 \\
     $\alpha_{aa}$ (\AA$^3$) & 21.5 & 34.506 & 28.847 & 22.072 & 25.774 \\
     $\alpha_{bb}$ (\AA$^3$) & 15.3 & 21.662 & 19.789 & 24.830 & 18.234 \\
     $\alpha_{cc}$ (\AA$^3$) & 10.2 & 12.123 & 11.292 & 11.103 & 10.192\\
     $\alpha_{ab}$ (\AA$^3$) & {0} & -1.209 & 0 & 0  & -0.419 \\
     $V_{aa} (1)$ (a.u.) & {$-1892.039/eQ$} & 12.327  &  7.34 & -0.30 &  8.3735 \\
     $V_{bb} (1)$ (a.u.) & {$978.816/eQ$} & -6.6802 & -1.73 &  5.88 & -2.7590 \\
     $V_{cc} (1)$ (a.u.) & {$913.222/eQ$} & -5.6472 & -5.61 & -5.57 & -5.6146 \\
     $V_{ab} (1)$ (a.u.) & {0} & 1.5092 & -7.85 &  8.84 & -7.1745 \\
     $V_{aa} (2)$ (a.u.) & {0} & 12.155  &  7.34 & -0.30 &  4.1656 \\
     $V_{bb} (2)$ (a.u.) & {0} & -6.4317 & -1.73 &  5.88 & -0.4784 \\
     $V_{cc} (2)$ (a.u.) & {0} & -5.7230 & -5.61 & -5.57 & -3.6872 \\
     $V_{ab} (2)$ (a.u.) & {0} & 0.57071 &  7.85 & -8.84 &  5.6800 \\
     \hline\hline
   \end{tabular}
   \caption{Rotational constants $A$, $B$, $C$, asymmetry parameter $\kappa$, elements of the
      polarizability tensors $\alpha_{ij}$ and electric field gradient tensors $V_{ij}(l)$ of the
      $l$th nucleus in the coordinate system defined by the principle axes of inertia. In the case
      of 1,3-BIB, $V_{ij}(1)$ and $V_{ij}(2)$ refer to the EFG of the iodine and bromine nucleus,
      respectively. For IB, the rotational constants from ref.~\onlinecite{Neill:JMolSpec269:21} and
      the polarizability tensor from ref.~\onlinecite{Poulsen:JCP121:783} were used. %
      \newline $^a$~In the calculations for IB we used experimentally determined quadrupole coupling
      constants~[\onlinecite[Table~6]{Neill:JMolSpec269:21}]. %
   }
   \label{tab:parameters}
\end{table*}
The rotational constants, asymmetry parameters $\kappa=(2B-A-C)/(A-C)$ and elements of the
polarizability and EFG tensors of the iodine and bromine nuclei in the principle axes of inertia frame are
listed in \autoref{tab:parameters}. Except for IB, the rotational constants were computed from the
equilibrium geometry obtained using density functional theory (DFT) with the B3LYP functional and
the def2-QZVPP basis set~\cite{Weigend:JCP119:12753,Weigend:PCCP7:3297}. The effective core
potential def2-ECP~\cite{Peterson:JCP119:11113} was used for the iodine atoms. The electric field
gradient and polarizability tensors of DIBN, 1,3-DIB and 1,2-DIB were calculated at the DFT/B3LYP
level of theory using the all-electron scalar relativistic Douglas-Kroll-Hess
Hamiltonian~\cite{Neese:JCP122:204107} with the DKH-def2-TZVP basis
set~\cite{Jorge:JCP130:064108,Campos:MolPhys111:167}. All electronic structure calculations were
carried out with the quantum-chemistry package ORCA~\cite{Neese:wircms2:73, Neese:wircms8:e1327}.
For IB, we used the experimental nuclear-quadrupole coupling constants~\cite{Neill:JMolSpec269:21}
$\chi_{aa}=-1892.039~\text{MHz}$, $\chi_{bb}=978.816~\text{MHz}$, $\chi_{cc}=913.222~\text{MHz}$
instead of a calculated electric field gradient and quadrupole moment.
The molecule-fixed frame ($x,y,z$) is
defined by the principle axes of inertia so that the $z$ axis is parallel to the most polarizable
axis or, in the case of DIBN, forms the smallest possible angle ($\approx5^\circ$) with it. The $z$
axis is thus chosen parallel to the $a$ axis for IB, DIBN, 1,3-DIB and 1,3-bromoiodobenzene and to
the $b$ axis for 1,2-DIB.

\section{Matrix representation of the Hamiltonian}
\label{sec:numerical_methods}
To obtain the eigenstates of $H_\text{mol}$, we first solved the time-independent Schr{\"o}dinger
equation for $H_\text{rot}$ yielding the rigid-rotor energy levels and eigenstates
$\ket{J_{K_aK_c},M_J}$. Here, $M_J$ is the eigenvalue of $\hat{J}_Z$ and $K_{a,c}$ are associated
with the projections $\hat{J}_{a,c}$ and are good quantum numbers in the prolate and oblate
symmetric top cases, respectively. The matrix representation of $H_\text{mol}$ was then constructed
in the coupled basis~\cite{Cook:AJP39:1433,Yachmenev:JCP147:141101} $\ket{F,J_{K_aK_c},I,M}$, $M$
being the eigenvalue of $\hat{F}_Z$. The field-free energies and eigenstates
$\ket{F,\tilde{J}_{\tilde{K}_a\tilde{K}_c},n,M}$ were computed by numerically solving the associated
time-independent Schr{\"o}dinger equation. Here, $n=1,2,...$ is an index labeling states with the
same $F$ and the approximate quantum numbers $\tilde{J}$, $\tilde{K}_a$ and $\tilde{K}_c$ were
assigned according to the coupled basis state $\ket{F,J_{K_aK_c},I,M}$ with the largest contribution
to the given eigenstate. For IB, the nuclear spin quantum number $I=I_1$ remains a good quantum
number. The matrix elements of the polarizability element $\alpha_{ZZ}$ are known
analytically~\cite{Cook:AJP39:1433} in the basis $\ket{F,J_{K_aK_c},I,M}$ and can be transformed
easily into the field-free eigenbasis~\cite{Yachmenev:JCP151:244118}.

\bibliography{string,cmi}

\onecolumngrid
\clearpage
\let\textCR\newline\listoffixmes
\end{document}